\documentclass[aps,prd,onecolumn,showpacs,groupedaddress]{revtex4-2}
\usepackage{graphicx}
\usepackage{dcolumn}
\usepackage{bm}
\usepackage{color}
\usepackage{subfigure}
\usepackage{longtable}
\usepackage{supertabular}
\usepackage{CJK}
\usepackage[T1]{fontenc}
\usepackage{epstopdf}
\usepackage{amsmath}
\usepackage{amssymb}
\usepackage{setspace}
\usepackage{multirow}
\usepackage{booktabs}
\usepackage[section]{placeins}
\usepackage{mathrsfs}
\usepackage{hyperref}
\usepackage{mnsymbol}
\usepackage[perpage]{footmisc}
\usepackage{footnote}

\newcommand{\Rmnum}[1]{\expandafter\@slowromancap\romannumeral #1@} 
\makeatother

\begin{document}
\title{Practicing carpe diem in the journey of studying physics\\
  \textemdash \small A brief review of the scientific contribution of Ru-Keng Su}

\author{Shaoyu Yin\textsuperscript{1}}
\author{Wei-Liang Qian\textsuperscript{2,3}}
\author{Ping Wang\textsuperscript{4}}
\author{Bin Wang\textsuperscript{3,5}}
\author{Rong-Gen Cai\textsuperscript{6,7,8}}\email[This article is dedicated to the memory of the late Professor Ru-Keng Su.]{}

\affiliation{$^{1}$ Institute for Theoretical Physics \& Cosmology, Zhejiang University of Technology, Hangzhou 310032, China}
\affiliation{$^{2}$ Escola de Engenharia de Lorena, Universidade de S\~ao Paulo, 12602-810, Lorena, SP, Brazil}
\affiliation{$^{3}$ Center for Gravitation and Cosmology, College of Physical Science and Technology, Yangzhou University, Yangzhou 225009, China}
\affiliation{$^{4}$ Institute of High Energy Physics, CAS, P. O. Box 918(4), Beijing 100049, China}
\affiliation{$^{5}$ School of Aeronautics and Astronautics, Shanghai Jiao Tong University, Shanghai 200240, China}
\affiliation{$^{6}$ CAS Key Laboratory of Theoretical Physics, Institute of Theoretical Physics, Chinese Academy of Sciences, Beijing 100190, China}
\affiliation{$^{7}$ School of Physical Sciences, University of Chinese Academy of Sciences, Beijing 100049, China}
\affiliation{$^{8}$ School of Fundamental Physics and Mathematical Sciences, Hangzhou Institute for Advanced Study, University of Chinese Academy of Sciences, Hangzhou 310024, China}

\begin{abstract}
We briefly review the scientific contributions of the late Prof. Ru-Keng Su in his academic life.
In the area of intermediate and high-energy nuclear physics, Su explored various topics in high-energy nuclear physics and particle physics, inclusively about the finite temperature field theory, effective models for nuclear and quark matter, soliton, and quasiparticle models, among others.
In gravity and cosmology, Su's research primarily embraces black hole thermodynamics, quasinormal modes, cosmological microwave background radiation, modified theories of gravity, and AdS/CFT correspondence and its applications.
Besides, many aspects of Su's distinguished impact on the Chinese academic physics community are discussed.
We also summarize the biographical and academic career of Su.
This article is an elaborated version of the memorial article that will be published in \href{https://www.mdpi.com/journal/symmetry}{\it symmetry}.

\end{abstract}

\date{Oct. 12th, 2022}

\maketitle

\section{Introduction}\label{sec1}

\begin{CJK}{UTF8}{min}
Ru-Keng Su (May 27, 1938 - June 3, 2022) was a highly respected Chinese theoretical physicist whose research interests span from the depth of the subatomic realm to the far reaches of the universe. 
During his academic career, Su made notable contributions to intermediate and high-energy nuclear physics, general relativity, and cosmology.
As one of the leading physicists in China, he has assisted in developing Chinese physical society to date.
This article elaborates on a few scientific results in memory of Su's contributions to the community.
\end{CJK}

The remainder of the article is organized as follows.
The following Sec.~\ref{section2} and~\ref{section3} are devoted to discussing Su's contributions, respectively, to the areas of nuclear and particle physics and general relativity. 
We give a brief account of relevant studies and, inclusively, those carried out in collaboration with his students and collaborators. 
The appendices consist of a brief biography and a summary of academic achievements of Su. 
As his disciples, we understand that it would be beneficial and educative for the followers to revisit his academic achievements as well as the personal life of our beloved supervisor. 
With the help of his family and friends, we tried to collect as much information as possible. 
Although we are aware that the writing may be subjective, as they are presented from our point of view, we sincerely hope it will help the readers appreciate how extraordinary Su's life has been. 

\begin{figure}[htb]
\centering
\includegraphics[width=0.4\textwidth]{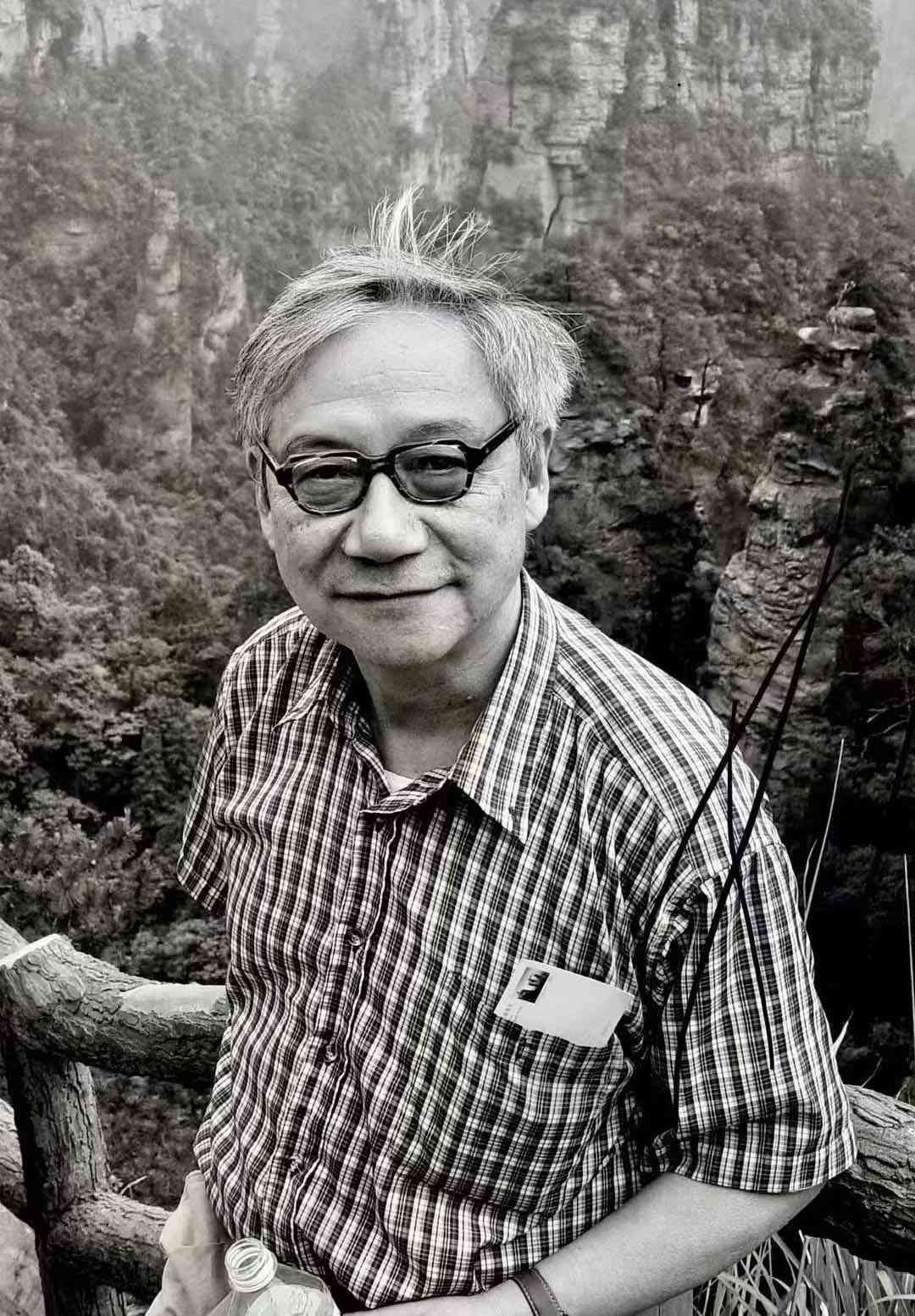}
\caption{Ru-Keng Su, in 2007 at Zhang Jiajie, Hunan.}
\label{portrait01}
\end{figure}

\section{Publications}

Overall, Prof. Su has published more than 200 academic papers in prestigious journals in physics, with a wide range of topics covering from the microscopic particles to the immense universe. 
His research interest can be categorized mainly into (1) nuclear and particle physics and (2) general relativity (GR) and cosmology.

\subsection{Nuclear and Particle Physics}\label{section2}

Su explored various topics in intermediate and high energy nuclear physics and particle physics, inclusively about the gauge fields~\cite{Yin:1980uq}, Higg mechanism~\cite{Ni:1980vc, Ni:1981mv}, soliton solutions and confinement~\cite{Su:1983bq, Su:1984gq, Su:1985mp, Su:1986ac, Brown:1986dg, Su:1987sw, Su:1989sf, Su:1989ah, Pan:1990vy, Su:1990fh, Liang:1990ky, Song:1991uv, Pan:1992sh, Su:1992wx}, and vacuum and fractional charge~\cite{Ni:1984jj, Ma:1985bj, Su:1987kq, Su:1988ta}. 
Notably, Su is well-known for his significant contributions to finite-temperature field theory~\cite{Su:1982un}, embracing both the real-time~\cite{Su:1983bq, Ni:1984er, Su:1984pyo, Su:1986ac, Su:1987kq, Chen:1988nh, Su:1989sf, Su:1989zz, Su:1990fh, Su:1991ni, Su:1992hy, Qian:1993pj, Qian:1994pp, Gao:1994wq, Su:1995zj, Gao:1995nh, Gao:1996fn, Gao:1996hx} and imaginary time~\cite{Su:1989ah, Pan:1990vy, Su:1990dy, Zheng:1990fv, Zheng:1990cx, Pan:1992sh, Su:1992np, Su:1993aa, Hu:1993rv} formalisms. Among others, these studies were primarily carried out with his students Song Gao, Yi-Jun Zhang, Zhi-Xing Qian {\it et al.}

In~\cite{Su:1983bq}, using generalized Bogoliubov transformation associated with the coherent state, a kink-soliton solution in $(1+1)$ dimensional real $\phi^4$ theory is derived.
The solution is interpreted as transitioning from a normal to the superfluidity state triggered by a spontaneous symmetry breaking of the vacuum state.
Subsequently, the spectra of the underlying system at zero and finite temperature are evaluated using the real-time Green function formalism.
The properties of the phase transition are elaborated, and as expected, the symmetry is restored at higher temperatures from which the critical temperature is derived.
In~\cite{Su:1989ah}, imaginary-time formalism is employed to explore the non-topological solitons proposed by Friedberg and Lee~\cite{friedberg-lee-01, friedberg-lee-02}, applied to the context of scalar soliton stars.
This type of soliton is triggered by the specific form of the effective potential and does not feature a topologically non-trivial profile.
By deriving the form of an effective Lagrangian at finite temperature, the gravitational field equations are derived and solved analytically under reasonable approximation.
The physical characteristics of the soliton star are evaluated and compared to those obtained under mean-field scenarios and/or at vanishing temperatures.

Su's main contributions to nuclear physics consist of studies primarily on the effective models for nuclear and quark matter.
They have explored topics such as Coulomb instability~\cite{Song:1991uv, Song:1993zz, Song:1993mn, Zhang:1999kt, Qian:2001dw}, the speed of sound~\cite{Su:1988zz}, hadrons in medium~\cite{Wang:1999vq}, thermodynamics~\cite{Qian:1993zz, Li:1999te, Wang:1999tq, Wang:2000um}, phase transition~\cite{Qian:2000fq, Qian:2002rp} of hadronic matter, and the quasiparticle models.

In~\cite{Song:1991uv}, the Coulomb instability in hot nuclei is explored in the context of the Skyrme model.
As a self-consistent mean-field approach comprised of a zero-range interaction with kinetic and density-dependent terms, the model is a viable tool in investigating the nuclear structure and low-energy dynamics, where collective correlations have been a priori built into the theoretical framework. 
Utilizing finite-temperature real-time Green's function, a detailed analysis is carried out focusing on the stability in the low-temperature and high-density regions of the system.
The stability's bound, the limit temperature, is found to be approximately proportional to the critical temperature. 
By comparing various models, it is also argued that the analysis discriminates between different approaches.
The liquid-gas phase transition in the hadronic matter as a two-component system was explored in~\cite{Qian:2000fq}.
It was found that the isospin degree of freedom plays a crucial role in the phase transition of the underlying system.
The phase diagram in the parameter space regarding isospin asymmetry and concentration is significantly modified when the interaction becomes isospin dependent. 
In particular, a ``cut-off'' feature is observed, and the critical temperature in the original system is replaced by a limit temperature.
The latter, subsequently, gives rise to various scenarios of phase transition regarding how the system is initially prepared. 
The hadron properties~\cite{Zhang:1997cr, Wang:1999pi, Zhang:2000ug, Zhang:2000vk, Wang:2001gv, Song:2001hz, Wang:2002aq, Wang:2002aq, Song:2003xe}, thermodynamics~\cite{Qian:2003ct, Qian:2004td}, and phase transition~\cite{Qian:2001dw, Yang:2003up, Yang:2005qz} of strange hadronic matter were investigated.
The studies were also performed for systems of quark degree of freedom to scrutinize the thermodynamics~\cite{Song:1995np, Song:1996mk, Zhang:2003ka, Wu:2005ty, Wu:2007je} of quark matter,
and thermodynamics~\cite{Zhang:2001ih, Zhang:2002fs, Zhang:2002ii, Zhang:2002qq, Zhang:2003fn} of strange quark matter.
The stability conditions for strangelet and strange matter were closely analyzed.

Su has also made noticeable contributions to various relevant aspects of the quasiparticle model, as well as the thermodynamic properties and consistency.
In~\cite{quasiparticle-qmdd-wang-01}, the role of an additional contribution to the thermal potential and its consequential effect on the strange quark matter were explored.
A series of studies regarding the quark mass density- and temperature-dependent (QMDTD) model was performed in~\cite{Zhang:2001ih, Zhang:2002fs, Zhang:2002qq, Zhang:2003ka}.
The temperature dependence of the stable radius of a strangelet was discussed in~\cite{Zhang:2001ih}.
The temperature dependence of the bag constant $B$ was explored and shown to cure the divergence that occurred at vanishing baryon density in the phase diagram for the bulk strange quark matter of the original QMDTD model~\cite{Zhang:2002fs}.
A systematic analysis regarding the stability of strangelet was performed in~\cite{Zhang:2002qq} in the framework of the QMDTD model.
It was observed that stable strangelets are more likely to be encountered in the region with a sizeable negative electric charge and significant strangeness.
The analysis was then extended to the dibaryon systems~\cite{Zhang:2003ka} regarding different decay channels, and the results were found in good agreement with those obtained by the chiral SU(3) quark model. 
The QMDTD setup was then applied to the context of Friedberg-Lee soliton bag~\cite{friedberg-lee-01, friedberg-lee-02, friedberg-lee-03} nonlinear coupled to the sigma~\cite{Wu:2005ty} as well as omega~\cite{Wu:2007je} mesons.
The model was further extended to investigate the properties of deconfinement~\cite{quasiparticle-qmdd-su-03, Wu:2007je} and nuclear matter~\cite{quasiparticle-qmdd-wu-05}.
As an alternative approach to address the thermodynamic consistency, an additional fictitious degree of freedom was introduced~\cite{Yin:2008mh, Yin:2008me} to elaborate a generalized version of the first law of thermodynamics.
These works were primarily carried out in collaboration with Hong-Qiu Song, as well as Su's students Zhi-Xin Qian, Ping Wang, Yun Zhang, Wei-Liang Qian, Li Yang, Chen Wu, and Shaoyu Yin.

\subsection{General Relativity and Cosmology}\label{section3}

In GR and cosmology, Prof. Su's research was carried out mainly with his students Rong-Gen Cai, Bin Wang, Cheng-Gang Shao, Li-Hui Xue, Da-Ping Du, Weigang Qiu, Jian-Yong Shen, Songbai Chen, Qiyuan Pan, Shaoyu Yin, Chang Feng, etc., along with other collaborators. 

Prof. Su's interest in GR and cosmology dates back to his early publications after the Cultural Revolution. 
In several papers published in Chinese and later in English, he discussed cosmological responses, gravitational radiation, cosmological microwave background radiation, the open universe model, Dirac's cosmological model and large number assumption~\cite{Su:1982ut}, Einstein-Dirac equation, wormholes, higher-order gravitational theory, among others. 
Though these publications primarily aimed to introduce the research frontier to more Chinese readers, the selection revealed his insight, fast response, and personal interest. 
Such work prepared him for innovative research in GR, and some topics had long been among his favorite ones, which he dwelled on throughout his research career.

Prof. Su's formal research on GR started at a regular pace in 1992 with two publications on black holes and wormholes~\cite{Pan:1992sh, Shen:1992zu}. Since then, his research has been carried out simultaneously on both sides of nuclear-particle physics and the GR. Soon there was a big boost owing to the excellent collaboration with Rong-Gen Cai, a doctoral candidate under the supervision of Prof. Su. 
They studied various properties of different types of black objects, including thermodynamics for dilaton black holes, neutral or charged~\cite{Cai:1993aa, Cai:1994nu, Cai:1994dn}, as well as black strings and p-branes~\cite{Cai:1994np}, phase transitions of black holes~\cite{Su:1994zz, Cai:1995wq}, stability of Cauchy horizons~\cite{Cai:1995ux}, statistical mechanics~\cite{Cai:1995fp} and Hawking radiation~\cite{Cai:1995ah} of BTZ black holes.

In this period, the most fruitful direction of their research is about the thermodynamics and statistical properties of black objects. Prof. Su and his collaborators successfully applied the Landau nonequilibrium fluctuation and phase transition theory in discussing the phase transition of various types of black holes. 
Specifically, the discussions of some divergent second moments helped to clarify the nature of phase transitions~\cite{Cai:1993aa, Su:1994zz, Cai:1994nu, Cai:1995wq}. They also showed particular interest in the dilaton black hole~\cite{Cai:1993aa, Su:1994zz, Cai:1994nu, Cai:1995wq}, which stems from string theory and has many interesting characters such as the secondary hair~\cite{Coleman:1992} and unusual thermal properties~\cite{Preskill:1991, Holzhey:1992}.

A second boost appeared when Bin Wang joined Prof. Su's group as a doctoral candidate and later as a permanent staff member. 
Then Prof. Su and Prof. Wang formed a minority group in the Department of Physics at Fudan University, where most of the teams focused on condensed matter, but their group remained the most productive one as judged by their academic output. 
Their study on black objects was continued and grew even more diversified and productive~\cite{Wang:1996fr, Wang:1996hp, Wang:1998wn, Wang:1998aa, Wang:1998kb, Wang:1998ke, Wang:1999gj, Wang:1999qc, Wang:1999sk, Wang:2000mv, Qiu:2001dg, Shen:2005nu, Mo:2006tb, Chen:2007jz, Chen:2007pu, Chen:2007ay, Yin:2010ouq, Yin:2010ix}, and numerical simulation of quasinormal modes had also been developed later~\cite{Xue:2002xa, Xue:2003vs, Shen:2003zm, Du:2004jt, Shao:2004ws, Shen:2006pa, Shen:2007xk, Pan:2011hj}. 

Meanwhile, their research topic extended to cosmology, with rich topics including the influnce of particle process on cosmological evolution~\cite{Su:1998vu}, relation between the Cardy formula of entropy and the Friedmann equation in the context of AdS/CFT duality for the (A)dS universes~\cite{Wang:2001bf} as well as the brane universes~\cite{Wang:2001bv}, quasi-normal modes of de-Sitter space-time~\cite{Qiu:2002cp}, tachyonic inflation~\cite{Wang:2002sz}, brane-world~\cite{Du:2003qm}, curvature of the universe based on supernova measurement~\cite{Wang:2004jf}, cosmological constant and dark sectors~\cite{Wang:2004nqa,Shen:2004ck,Huang:2005zf,Huang:2006er,Feng:2007wn,Feng:2008fx}, cosmic microwave background radiation~\cite{Huang:2005re}, modified gravity such as using the Dirac cosmology, which he favored all along, to explain the accelerated expansion of contemprary universe~\cite{Shao:2005mg} and the $1/R$ gravity in the solar system tests~\cite{Shao:2005wt}, thermal effetc in inflation~\cite{Yin:2008ku}, entanglement entropy in holographic models~\cite{Cai:2013oma}, etc. 
It can be inferred from the publications' titles that the scope of the research was not limited to a thin branch. 
Generally speaking, a visible trend in their research direction is to expand from the purely mathematical study, such as the stability and geometry of black hole horizon in various dimensions of space-time~\cite{Wang:1996fr, Wang:1996hp, Wang:1998wn}, to topics more related to actual observations. 
This revealed their crucial value in keeping pace with the fast progress of cosmology and positively and vigorously participating the international competition. 
They did well, and their group became one of China's most active and prestigious research teams on GR and cosmology.

The collaboration between Su and Wang significantly contributed to the field of black hole quasi-normal modes and the constraints on the cosmological model. 
Among dozens of articles published in high-impact journals, several most frequently cited works include testing the viability of the interacting holographic dark energy model\cite{Feng:2007wn}, where they utilized observational constraints combining the latest gold type Ia supernova samples, Wilkinson Microwave Anisotropy Probe observations, the baryon acoustic oscillation measurement, $H(z)$ and lookback time measurements, the joint statistical analysis provided the state-of-the-art testing of the interacting holographic dark energy model contemporarily. They have also studied with a similar approach the constraints on the dark energy from the holographic connection the small $l$ CMB suppression~\cite{Shen:2004ck}, on dark energy from holography~\cite{Wang:2004nqa}, on the dark energy and dark matter mutual coupling~\cite{Feng:2008fx}, from which it can be seen how Prof. Su's group tried to catch up the academic frontier and to compete with the world-leading research teams.

Prof. Su's research on GR had some prominent characteristics. 
First of all, taking advantage of his expertise in nuclear and high-energy physics, he could make cross-field studies combining topics in GR with nuclear and particle physics or utilizing the thermal field theory as a handy tool, such as in the study of the scalar wormhole at finite temperature~\cite{Pan:1995yr} and astrophysics of compact objects involving both GR and nuclear/particle models~\cite{Pan:1992sh, Shen:2005vh, Yin:2010ouq}. 
Secondly, he encouraged students' initiative and would like the abled students to choose their favorite topics rather than simply make an assignment for them.
For instance, when Cheng-Gang Shao worked as a postdoc in Su's group, he studied gravitational test~\cite{Shao:2005wt}, slightly deviated from the mainstream of the group's research; while Li-Hui Xue, a student with excellent programming skills contributed significantly in developing the numerical codes. 
As a matter of fact, many new research branches in Su's group were initialized by some energetic students. 
However, it is worth emphasizing that, in all the projects, Prof. Su always keeps pace with the progress and makes necessary guidance with excellent insight. 
In Prof. Su's group, there was always an active and democratic atmosphere and open for any discussion. One of his slogans is ``I only worry about those of my students who do not dispute against me''.

\section{Other Scientific Writings}

Prof. Su authored several highly appreciated and widely used textbooks, including ``Quantum Mechanics'' (first edition in 1997 by Fudan University Press, second edition in 2002 by Higher Education Press), 
``Statistical Physics'' (first edition in 1990 by Fudan University Press, second edition in 2004 by Higher Education Press), 
``Advanced Quantum Mechanics'' (English version, co-authored with Bin Wang in 2004 by Fudan University Press), and 
``Challenges in Physics - Selected Frontiers and Basic Topics of Physics'' (in 1991 by Liaoning Education Press). 
He also translated A. Messiah's ``Quantum Mechanics'' (Volume 1, in collaboration with Jiayong Tang, published in 1986 by Science Press).

\appendix
\section{Life, Education, and Academic career}\label{AppSec1}

Born on May 27, 1938,  Prof. Ru-Keng Su was a native of Shunde, a prosperous town close to the Pearl (Zhujiang) River delta in the Guangdong Province of China. 
Though he rarely talked about his childhood, it can be inferred that he was born into a well-off family, as he suffered during the Cultural Revolution due to his parentage. 
This explains the excellent education he received and his exquisite taste in literature and classical Chinese poems.
At eighteen, he was admitted to Peking University, the top university in China, to study Physics. 
Subsequently, he left the southern border of China and traveled to the capital in the far north.
The journey was uneasy for him at that time, which took several days. 
He told us more than once very vividly how he took the boat to cross the Yangtze river at Wuhan when the famous bridge crossing the river had not been built yet and experienced distinct folk customs in Central China.

Though harassed by continual political movements and limited food rationing during that special period, Mr. Su won top-notch scores in his undergraduate studies, despite the disturbance of coercive collective labor on farmland in the daytime and volunteer work at night.
Specially, Mr. Su had been working extensively to assist Prof. Zhuxi Wang in composing a booklet of Concise 10-digit Logarithmic Tables as an offering to the first-decade celebration of P. R. China. 
He used a rolling calculator, a state-of-the-art facility in China then, while Prof. Wang double-checked his results using two abacuses.
His talent was recognized, and when he graduated in 1960, he was chosen to work at Fudan University, a prestigious university in Shanghai. 
Since then, he started his life there and contributed all his passion to teaching and research in the Department of Physics at Fudan University for more than half-century. 
He told us that on his first night in Shanghai, he wandered alone in the university stadium under the bright mid-autumn moon, hungry and pondering about his possible future at Fudan. 

Prof. Su felt fortunate that he was assigned to the division of theoretical physics, where he joined the research group of senior professor Shixun Zhou.
Prof. Su enthusiastically plucked himself into heavy teaching and research tasks.
Prof. Zhou composed a concise textbook on quantum mechanics, which became popular in numerous Chinese universities in the 1960s. 
This inspired Prof. Su to enrich the contents, notably more modern progress, and elaborate his renowned {\it Quantum Mechanics}, which has become one of China's most widely adopted textbooks.
Even during the difficult period, Su still managed to be active in research for as long as possible, which brought him trouble and punishment. 
Nevertheless, his optimism allowed him to be determined in his research and pursue a valuable life, even if he was persecuted and forced to take on unpleasant chores.
After the 70s, his effort was rewarded, demonstrated by the burst of publications.
He was promoted to associate professor in 1982, then to full professor in 1987, owing to his distinctive achievements in teaching and research. Since 1987, Prof. Su has started to serve as a Ph.D. supervisor. 

Encouraged by Prof. Chen-Ning Yang, Fudan University established a local research team on nuclear physics.
The team was led by Prof. Fujia Yang, whose members include, among others, Prof. Chaohao Gu and Prof. Daqian Li. 
As an active team member, Prof. Su's expertise and devotion were highly appreciated. 
Subsequently, he was invited by Prof. Yang and visited the nuclear physics research group at Stony Brook, New York State University, three times between 1984 and 1990. 
During his stay in the United States, Prof. Su also visited the Institute of Nuclear Physics at the University of Washington in Seattle in 1985, where he collaborated with E.M. Henley.
Besides, he visited the Department of Physics and Astronomy, University of Kentucky, from September 1989 to February 1990. 
Prof. Su loved to tell the following story, especially to those heading for the States to pursue their studies.
Once, he had to transfer in Chicago on a journey from Seattle to New York. 
Having heard of violent crimes in the city, he was pretty worried about being robbed of his hard-earned salary carrying with himself. 
Taking advice from some Chinese fellows, he decided to wear a pair of sunglasses and a black overcoat with his hands stuck in the pockets, forming the shape of a gun. 
Having camouflaged himself in such an over-the-top stereotype of the local Chinese gangster and strode defiantly through the crowd in public, Prof. Su arrived at his destination safely and happily, with himself and the cash intact. 

In the 1990s, he worked at the City University of Hong Kong for a couple of years, where he made some good friends, including Prof. Jiju Xiao, who used Prof. Su's {\it Quantum Mechanics} as a textbook in his lessons.
It was the first Chinese physics textbook adopted in the university, extending its readership beyond Mainland China.
At the beginning of the 1990s, he was invited by Prof. Tsung-Dao Lee and became a member of the advisory committee at the Chinese Center of Advanced Science and Technology. 
As a reward for his academic activity and support for the center, as well as a personal expression of friendship, for many years, Prof. Su received greeting cards or books with Prof. Lee's paintings as New Year gifts.

\section{Notable status and awards}\label{AppSec2}

Given his academic achievement, Prof. Su had been elected as the vice-chairman of the Chinese Society of High Energy Physics and a member of the Senate in the Division of Mathematics and Physics of the National Natural Science Foundation of China (NSFC). 
Prof. Su was also an active member of the Center of Theoretical Nuclear Physics in the National Laboratory of Heavy Ion Collisions in Lanzhou. 
In his public services, his integrity and insight were highly appreciated by his colleagues and peers. 
Many junior scientists are still grateful for the unbiased and unselfish support they received from him.

Su received three times the second prize for Scientific and Technological Development issued by the Ministry of Education.
He was awarded in 1988 for ``Vacuum stability, spontaneous symmetry breaking, and thermal field theory'' (in collaboration with Guang-Jiong Ni), 
in 1992 by ``Theoretical research on phase transition in nuclear matter at finite temperature and density'', and in 1996 by ``Temperature field theory and its application in nuclear physics and astrophysics''.
In 1999, he won the second prize in the prestigious Natural Science Award issued by the Chinese Academy of Science for ``Critical phenomena in nuclear systems and the effect of many-body correlation'' (in collaboration with Hongqiu Song).
In 2003, he received the second prize in the Shanghai Science and Technology Achievement Award for ``Theoretical research of holographic principle in black hole physics and cosmology'' (in collaboration with Bin Wang).

He has served on national boards of physics and astronomy foundations and played a pivotal role in promoting the development of Chinese physics society. 
In particular, he consistently served as a member of the core evaluation committee of the National Natural Science Foundation of China (NSFC) and the academic advisory committee of the China Center of Advanced Science and Technology (CCAST).

\section{Miscellaneous contributions}\label{AppSec3}

Su's interest in physics is not limited to the above topics in theoretical physics. 
He also significantly contributed to science popularization in China, fulfilling the vital task of making scientific knowledge understandable and accessible to the lay public.
Moreover, Su had also deeply indulged himself in other fundamental problems in theoretical physics.
Relevant topics include the time arrow, negative temperature, hidden variables of quantum mechanics, neutrino mass, the essence of light, coherent states, and the Klein paradox, among others. 
He was not only among the first few Chinese physicists working on finite-temperature field theory but also played a vital role in promoting its usage by fellow Chinese physicists. 

Besides scholastic publications, Su is well-known for his talent in classical Chinese poems in the Chinese high-energy physics community. 
For a specific period in the past, Su and his peers were accustomed to sending correspondence in the form of poems to express their feelings about research and life.
Su later showed those pieces to the students who attended his lectures. 
He had also published two reviews in the ``Journal of Football'' which, according to some readers, demonstrated better logic than some experts of the sport.

\section{Teaching and orientations}\label{AppSec4}

Prof. Su had taught at Fudan University for more than half a century. 
His lectures covered all major physics courses and contributed significantly to the curriculum reform in Fudan. 
For undergraduate students, he lectured ``quantum mechanics'', ``thermodynamics'', ``statistical physics'', ``classical mechanics'', ``modern physics'', ``methods of mathematical physics'', 
among others.
For graduate students, he had given ``advanced quantum mechanics'', ``advanced statistical physics'', ``many body theory'', ``frontiers in nuclear and particle physics'', ``quantum field theory'', ``thermal field theory'', 
``general relativity and cosmology'', ``soliton and instanton'', ``quantum field theory in curved spacetime''. 
In 1999, his textbook ``quantum mechanics'' was awarded the first prize in Shanghai Excellent Teaching Materials for Universities.
He also played a major role in tutoring students to prepare for CUSPEA, which contributed significantly to the excellent scores won by the students from Fudan.

Su was highly acclaimed for his humorous, passionate, and modern teaching methodology. 
He was a prominent educator who cast a remarkable constructive impact on his students. 
Students overwhelmingly adored him for his profound insights and clear explanations and rated him as the most popular teacher at Fudan for many successive years.
Over the decades, tens of thousands of students have listened to his lectures in person, while countless others have studied following his textbooks and online video records.

\begin{figure}[htb]
\centering
\includegraphics[width=0.7\textwidth]{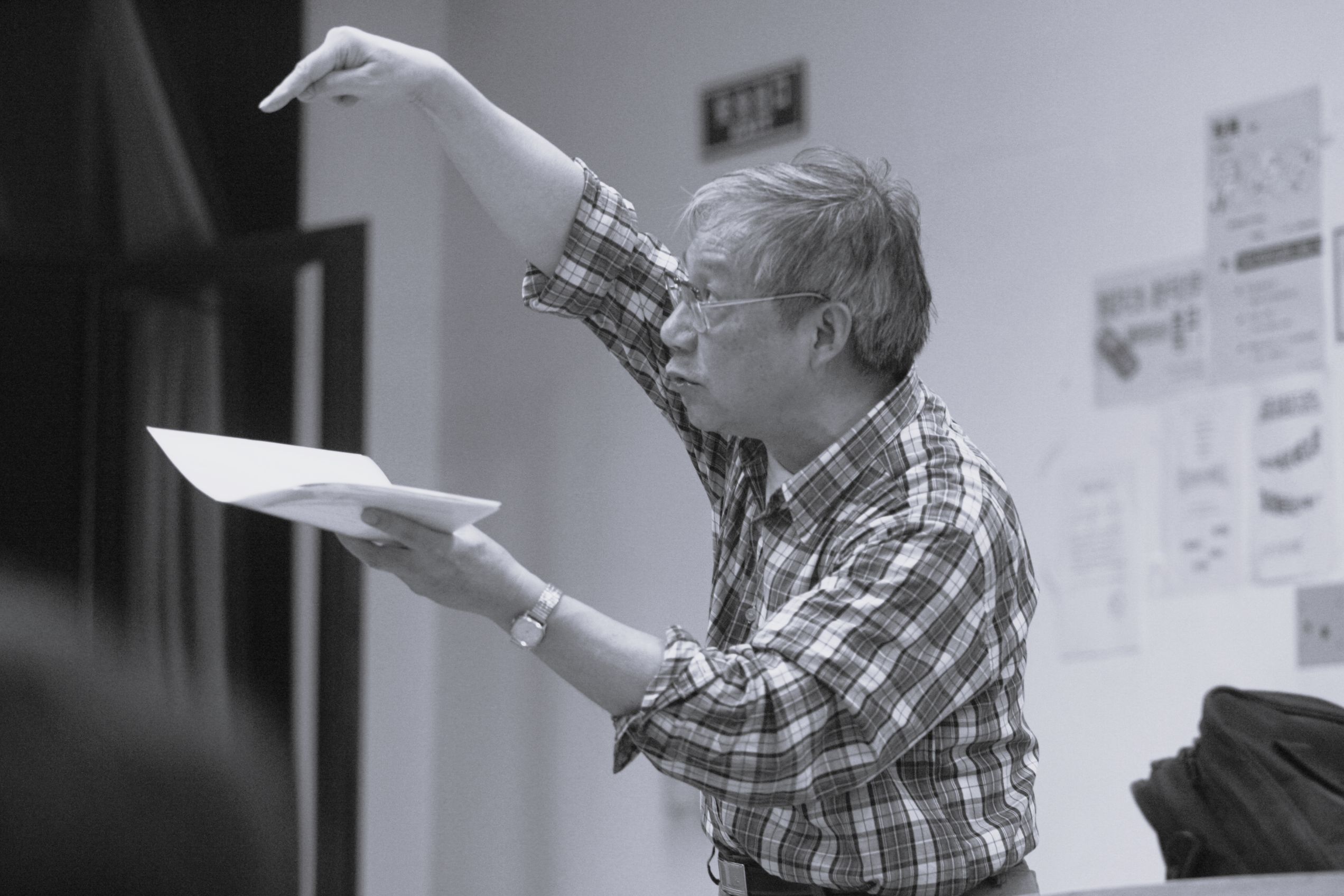}
\caption{Prof. Ru-Keng Su was well-known for his patent energy and expressive body language in the classroom.}
\label{portrait01}
\end{figure}

In 1993, Su was awarded the second prize in the Shanghai Excellent Teaching Achievement Award for ``constantly reforming teaching content and methods to cultivate potential talents''.
In 1999, he won the Baosteel Excellent Teacher Award.
In 2001, he won the second prize in the Shanghai Excellent Education Achievement Award for ``merging the frontier of science into didactic materials and profound development of the pedagogy in teaching quantum mechanics''.
In 2003, he won the Shanghai Teaching Master Award.
As a part of the unique collection of colleague physics materials (in collaboration with Qimin Jia and Guangjiong Ni), his textbook won the first prize in Shanghai Excellent Education Achievement Award and the second prize in National Teaching Achievement Award; Soil" won the second prize of Shanghai Teaching Achievement Award. 
In 2004, ``quantum mechanics'' was selected as a national premier course for its excellency.
In 2016, the Ministry of Education included it as part of the first batch of ``Public Courses of National Excellent Resource''. 
The corresponding online course accumulated at least several hundred thousand views. 
Many audiences claimed that Su's energetic and charming teaching style inspired their interest, helped them to understand esoteric quantum mechanics, and even to pass the graduate school entrance exam.

Su cultivated eleven Ph.D. and dozens of masters and supervised hundreds of undergraduate students. 
The Ph.D. students are Rong-Gen Cai (1995), Song Gao (1995), Yi-Jun Zhang (1997), Bin Wang (1998), Ping Wang (1999), Yun Zhang (2003), Wei-Liang Qian (2003), Wei-Gang Qiu (2005), Jian-Yong Shen (2008), Chen Wu (2009), and Shaoyu Yin (2010).
There also were three postdoctoral fellows: Shuqian Ying (1993-1997), Cheng-Gang Shao (2004-2006), and Songbai Chen (2006-2008). 
These students who have benefited from Su's teaching are engaged in various trades worldwide, and many are working at the front line of scientific research and education. 
Su's knowledge and spirit will continue to be passed on and carried forward from generation to generation.

\bibliographystyle{h-physrev}
\bibliography{references_su_citeall, references_qian}

\end{document}